\documentclass[a4paper,twocolumn,11pt,accepted=2025-04-30]{quantumarticle}
\pdfoutput=1
\usepackage[utf8]{inputenc}
\usepackage[english]{babel}
\usepackage[T1]{fontenc}
\usepackage{amsmath,amssymb}
\usepackage{amsthm,xcolor}
\usepackage{subfig}
\usepackage{hyperref}

\usepackage{tikz}
\usepackage{lipsum}

\usepackage{float}
\usepackage{graphicx}
\usepackage{braket}
\usepackage{mathtools}
\usepackage{dsfont}
\usepackage[capitalise]{cleveref}
\usepackage[numbers,sort&compress]{natbib}
\usepackage{makecell}

\newtheorem{theorem}{Theorem}

\newtheorem{observation}{Observation}

\def\tr {\mathrm{Tr}}

\begin{document}

\title{An Effective Way to Determine the Separability of Quantum State}

\author{Ma-Cheng Yang}
% \email{yangmacheng@ucas.ac.cn}
\affiliation{School of Physical Sciences, University of Chinese Academy of Sciences, YuQuan Road 19A, Beijing 100049, China}
\orcid{0000-0003-4122-4310}

\author{Cong-Feng Qiao}
\email{qiaocf@ucas.ac.cn}
% \homepage{http://quantum-journal.org}
\orcid{0000-0002-9174-7307}
% \thanks{You can use the \texttt{\textbackslash{}email}, \texttt{\textbackslash{}homepage}, and \texttt{\textbackslash{}thanks} commands to add additional information for the preceding \texttt{\textbackslash{}author}. If applicable, this can also be used to indicate that a work has previously been published in conference proceedings.}
\affiliation{School of Physical Sciences, University of Chinese Academy of Sciences, YuQuan Road 19A, Beijing 100049, China}
\affiliation{Key Laboratory of Vacuum Physics, CAS, YuQuan Road 19A, Beijing 100049, China}
% \author{Marcus Huber}
% \affiliation{Institute for Quantum Optics \& Quantum Information (IQOQI), Austrian Academy of Sciences, Boltzmanngasse 3, Vienna A-1090, Austria}
% \orcid{0000-0003-1985-4623}
% \author{Cassandra Granade}
% \affiliation{Microsoft Research, Quantum Architectures and Computation Group, Redmond, WA 98052, USA}
% \author{Johannes Jakob Meyer}
% \affiliation{Dahlem Center for Complex Quantum Systems, Freie Universität Berlin, 14195 Berlin, Germany}
% \orcid{0000-0003-1533-8015}
% \author{Victor V. Albert}
% \affiliation{Institute for Quantum Information and Matter \& Walter Burke Institute for Theoretical Physics, Caltech, Pasadena, CA 91125, USA}
% \orcid{0000-0002-0335-9508}
\maketitle

\begin{abstract}
We propose in this work a practical approach to address the longstanding and challenging problem of quantum separability, leveraging the correlation matrices of generic observables. General separability conditions are obtained by dint of constructing the measurement-induced Bloch space, which in essence come from the intrinsic constraints in the space of quantum state. The novel approach can not only reproduce various established entanglement criteria, it may as well brings about some new results, possessing obvious advantages for certain bound entangled states and the high dimensional Werner states. Moreover, it is found that criteria obtained in our approach can be directly transformed into entanglement witness operators.
\end{abstract}

\section{Introduction}\label{sec:intro}
\noindent
Quantum entanglement captures the most enigmatic and deepest insights of quantum theory, and has a broad connections with other branches of physics \cite{vedral14}. In quantum information theory, distinguishing entangled states from separable states appears to be a fundamental and challenging issue, commonly referred to as the separability problem \cite{horodecki09,guhne09}. It is widely recognized that the separability problem becomes NP-hard as the dimension increases, even in bipartite systems \cite{gurvits04}. Therefore, developing a practical and experimentally feasible separability criterion is crucial for establishing a comprehensive entanglement theory \cite{horodecki09,guhne09}. In past decades, a multitude of approaches have been proposed to address this issue, including positive map theory and entanglement witnesses \cite{horodecki96-TXC1O,peres96,horodecki01}, spin squeezing inequalities \cite{sorensen01,toth07,toth09_spin,ma11}, the computable cross norm or realignment (CCNR) criterion \cite{rudolph00,chen03,rudolph05}, correlation matrix method \cite{vicente07,de08-rQF69,li18,shang18,sarbicki20}, local uncertainty relation (LUR) criteria \cite{duan00,hofmann03,guhne04-frAQB,zhang07,zhang10,schwonnek17}, covariance matrix \cite{guhne07,gittsovich08}, quantum Fisher information (QFI) method \cite{hyllus12,li13,ren21,liu24_qfient}, moment method \cite{shchukin05,elben20,neven21-moment,yu21}, etc.

In this paper, we propose a comprehensive framework to establish a separability criterion for finite-dimensional quantum systems using the correlation matrix. To achieve this, we introduce a measurement-induced generalized Bloch space for arbitrary observables, which enables us to formulate a generic separability criterion. A critical observation is that the correlation matrix of separable states is constrained by the scale parameter of the generalized Bloch space. The versatility and strength of this approach are evident from its ability to reproduce several established criteria, such as de Vicente's correlation matrix criterion \cite{vicente07}, the CCNR criterion \cite{chen03,rudolph05}, the correlation matrix criterion based on symmetric informationally complete positive operator-valued measures (SIC-POVM) \cite{shang18} and Sarbicki \emph{et al.}’s criterion \cite{sarbicki20}, while the potential to yield some other even stronger results. Additionally, this framework provides a streamlined method for constructing entanglement witness operators using generic observables. We provide examples demonstrating the effectiveness of detecting bound entangled states and the high-dimensional Werner states.

\section{Bloch representation of observables and density matrix}

Before proceeding, we introduce the necessary notations. An observable is denoted by a capital letter, such as $A$, while a boldface capital letter $\boldsymbol{A}$ represents an $m$-tuple of observables, i.e., $\boldsymbol{A} = (A_1, A_2, \cdots)$. The density matrix $\rho$ of a $d$-level quantum system is expressed in the Bloch representation, specifically \cite{fano57,harriman78,hioe81,byrd03,kimura03,li-qiao15}
\begin{align}
\rho = \frac{1}{d}\mathds{1} + \frac{1}{2}\vec{r}\cdot\vec{\pi} \; .
\label{eq:bloch_rep}
\end{align}
Here, $\vec{\pi}$ are $d^2-1$ generators of $\mathfrak{su}(d)$ Li algebra. Due to the orthogonality relation of the generators $\tr[\pi_{\mu}\pi_{\nu}]=2\delta_{\mu\nu}$, we have $r_{\mu}=\tr[\rho \pi_{\mu}]$. The set of the Bloch vectors $\vec{r}$ constitutes the so-called Bloch space. The positive semidefiniteness of density matrix constrains the size of Bloch space, i.e. $|\vec{r}|\leq \sqrt{2(d-1)/d}$. Similarly, the Bloch representation of an observable $A_{\mu}$ reads
\begin{align}
A_{\mu} = & \frac{t_{\mu}}{d}\mathds{1} + \vec{a}_{\mu}\cdot\vec{\pi} \nonumber\\
= & \frac{t_{\mu}}{\sqrt{2d}}\sqrt{\frac{2}{d}}\ \mathds{1} + \vec{a}_{\mu}\cdot\vec{\pi} \; ,
\end{align}
where $t_{\mu} =\tr[A_{\mu}]$ and $a_{\mu\nu} = (\vec{a}_{\mu})_{\nu} = \tr[A_{\mu}\pi_{\nu}]/2$. And then, an $m$-tuple $\boldsymbol{A}$ of observables can be parameterized as a real matrix $M$ in Bloch representation
\begin{align}
\boldsymbol{A} \rightarrow M_A \equiv \left[
\begin{matrix}
\frac{t_{1}}{\sqrt{2d}} & \frac{t_{2}}{\sqrt{2d}} & \cdots \\
\vec{a}_{1} & \vec{a}_{2} & \cdots
\end{matrix}
\right] \; .
\label{eq:measure_vec_bloch}
\end{align}
Note, here $\vec{a}_{\mu}$ are column vectors.

\section{Correlation matrix of separable state and measurement-induced Bloch space}

Quantum entanglement exhibits inherent non-classical correlations in many-body systems, even when the particles are far apart. This phenomenon was initially recognized by Einstein, Podolsky, and Rosen \cite{einstein35}, as well as by Schrödinger \cite{schrodinger35}. Based on correlation functions, Bell inequalities provide a practical method to test entanglement for the first time \cite{bell64}. Among the family of Bell inequalities, the Clauser-Horne-Shimony-Holt (CHSH) inequality \cite{clauser69}
\begin{align}
\left|\! \braket{A\otimes B}\!\! - \!\!\braket{A\otimes B'}\!\! + \!\!\braket{A'\otimes B}\!\! + \!\!\braket{A'\otimes B'} \!\right|\! \leq \!2 
\label{eq:chsh}
\end{align}
is perhaps the most prominent. Here, $A(A')$ and $B(B')$ are pairs of dichotomic observables with eigenvalues of $\pm 1$ measured by the parties $A$ and $B$, respectively. The violation of CHSH inequality and the predictions of quantum theory were first confirmed experimentally by \cite{freedman72,aspect82,aspect82-frAQB}, then later more convincingly by some loophole-free tests \cite{rowe01,hensen15,handsteiner17,collaboration18}. Although the violation of the CHSH inequality certifies entanglement and excludes the possibility of a local hidden variable description for correlated systems, Bell inequalities inherently fail to detect all entangled states \cite{werner89}. Therefore, to fully explore and utilize the nature of entanglement, it is essential to establish even stronger criteria for determining separable states.

To this end, we define the following correlation matrix entries for arbitrary $m$-tuples $\boldsymbol{A}$ and $\boldsymbol{B}$
\begin{align}
\mathcal{C}_{\mu\nu} \equiv \braket{A_{\mu}\otimes B_{\nu}} \; .
\label{eq:correlation_matrix}
\end{align}
A state is separable iff it is a convex combination of product states, i.e. $\rho_{\mathrm{sep}}=\sum_{i}q_{i}\rho_{i}^{A}\otimes \rho_{i}^{B}$, where $\{\rho_{i}^{A}\}$ and $\{\rho_{i}^{B}\}$ are two pure subsystems \cite{werner89}. Hence the correlation matrix of a separable state can be written as
\begin{align}
\mathcal{C} = \sum_{i}q_{i}\left(
\begin{matrix}
\braket{A_{1}}_{i} \\
\braket{A_{2}}_{i} \\
\vdots
\end{matrix}
\right)\left(
\begin{matrix}
\braket{B_{1}}_{i}, \braket{B_{2}}_{i}, \cdots
\end{matrix}
\right) \; ,
\label{eq:sep_corre_matrix}
\end{align}
where $\braket{A_{\mu}}_{i}=\tr[\rho_{i}^{A}A_{\mu}]$ and $\braket{B_{\nu}}_{i}=\tr[\rho_{i}^{B}B_{\nu}]$. \cref{eq:sep_corre_matrix} inspires us to define a measurement-induced Bloch vector (MIBV) $\vec{\alpha}$ for $\boldsymbol{A}$, given by
\begin{align}
\alpha_{\mu} \equiv \tr[\rho \bar{A}_{\mu}] = \vec{r}\cdot\vec{a}_{\mu} \;
\label{eq:gene_bloch_vec}
\end{align}
where $\bar{A}_{\mu}=\vec{a}_{\mu}\cdot\vec{\pi}$ denotes the traceless part of the observable $A_{\mu}$. Various MIBV $\vec{\alpha}$ span the corresponding measurement-induced Bloch space (MIBS), denoted as $\mathcal{B}(\boldsymbol{A})$, which is defined as
\begin{align}
\mathcal{B}(\!\boldsymbol{A}\!)\! = \!\{\vec{\alpha}\!=\!(\braket{\!\bar{A}_1\!}_\rho,\!\cdots,\!\braket{\!\bar{A}_m\!}_\rho)^{\mathrm{T}}\!|\rho\in \!\mathcal{S}(\mathcal{H})\!\} \; .
\end{align}
Here, $\mathcal{S}(\mathcal{H})$ is the set of density matrices. It is noteworthy that $\mathcal{B}(\boldsymbol{A})$ is also referred to as the joint numerical range (JNR) of observables $\bar{A}_1,\cdots,\bar{A}_m$ \cite{szymanski18,szyma20}, which has been extensively studied in the context of uncertainty relations, quantum entanglement, and more \cite{schwonnek17,xu24}. In particular, the quantity
\begin{align}
\beta \equiv \sup_{\rho}\sum_{k=1}^{m}\braket{\bar{A}_k}_{\rho}^2 = \max_{\vec{x}\in\mathcal{B}(\boldsymbol{A})}|\vec{x}|^2
\end{align}
is a critical parameter of JNR or MIBS, closely related to state-independent uncertainty relation \cite{de23,moran24}. In the following sections, we will establish an explicit relationship between $\beta$ and the quantum separability problem. The correlation matrix of a separable state can be reformulated using MIBVs, as follows:
\begin{align}
\mathcal{C} = \sum_{i}q_{i}(\vec{t}^{A}/d_{A}+\vec{\alpha}_{i})(\vec{t}^{B}/d_{B}+\vec{\beta}_{i})^{\mathrm{T}} \; .
\end{align}
Here, $\vec{t}^{A}$ and $\vec{t}^{B}$ are column vectors consisting of the traces of measurements $\boldsymbol{A}$ and $\boldsymbol{B}$, respectively; $\vec{\alpha}_{i}$ and $\vec{\beta}_{i}$ are MIBVs of measurements $\boldsymbol{A}$ and $\boldsymbol{B}$; and $\mathrm{T}$ denotes the transpose of a given matrix.

\section{The separability criteria}

With this, we are now ready to present the main results of this work, with proof given in the \cref{appen:proof_thm_sep}:
\begin{theorem}\label{thm:sep_cri}
Let $\mathcal{C}$ be a correlation matrix for arbitrary $\boldsymbol{A}$ and $\boldsymbol{B}$ with $\sum_{\mu} t_{\mu}^{A}\vec{a}_{\mu}=\sum_{\mu} t_{\mu}^{B}\vec{b}_{\mu}=0$, if the corresponding state $\rho$ is separable, then
\begin{align}
\|\mathcal{C}\|_{\operatorname{tr}} \leq \kappa_{A}\kappa_{B} \; .
\end{align}
Otherwise, $\rho$ will be an entangled state. Here, $\|X\|_{\operatorname{tr}}=\tr\left[\sqrt{X^{\dagger}X}\right]$ is the trace norm of matrix $X$; $\kappa_{A}=\sqrt{(\vec{t}^{\,A})^2/d_{A}^2+\beta_A}$ with $\beta_A=\sup_{\rho}\sum_{k=1}^{m}\braket{A_k}_{\rho}^2$, and similarly for $\kappa_{B}$.  
\end{theorem}
Note, \cref{thm:sep_cri} provides an alternative approach to determining the separability of quantum states, different from the covariance matrix (CM) criteria \cite{guhne07, gittsovich08}. The CM criteria make use of the concavity and the positive definitiveness of CM, while \cref{thm:sep_cri} stems from the geometry property of the MIBS, highlighting the independence of the covariance matrix and correlation matrix methods. Next, we illustrate how \cref{thm:sep_cri} works. For a bipartite system, we have the following Bloch representation
\begin{widetext}
\begin{align}
\rho &= \frac{1}{4}\sum_{\mu,\nu}\chi_{\mu\nu}\Pi_{\mu}^{A}\otimes\Pi_{\nu}^{B} \\
&= \frac{1}{d_{A}d_{B}}\mathds{1}\otimes\mathds{1} + \frac{1}{2d_{B}}\vec{a}\cdot\vec{\pi}^{A}\otimes\mathds{1} + \frac{1}{2d_{A}}\mathds{1}\otimes\vec{b}\cdot\vec{\pi}^{B} + \frac{1}{4}\sum_{\mu,\nu}\mathcal{T}_{\mu\nu}\pi_{\mu}^{A}\otimes\pi_{\nu}^{B} \, .
\label{eq:bipartite_state}
\end{align} 
\end{widetext}
Here, $\Pi_{0}^{X}=\sqrt{\frac{2}{d_{X}}}\mathds{1}$; $\{\Pi_{\mu}^{X}=\pi_{\mu}^{X}\}_{\mu=1}^{d^2-1}$ are the generators of the $\mathfrak{su}(d_{X})$ Lie algebra for $X=A,B$; coefficient matrix is given by $\chi_{\mu\nu}=\tr[\rho(\Pi_{\mu}^{A}\otimes\Pi_{\nu}^{B})]$;
$a_{\mu}=\tr[\rho \pi_{\mu}^{A}\otimes\mathds{1}]$; $b_{\mu}=\tr[\rho \mathds{1}\otimes\pi_{\mu}^{B}]$; $\mathcal{T}_{\mu\nu}=\tr[\rho \pi_{\mu}^{A}\otimes\pi_{\nu}^{B}]$; and $\chi_{00}=\frac{2}{\sqrt{d_{A}d_{B}}}$ since $\tr[\rho]=1$. The correlation matrix can then be expressed as
\begin{align}
&\mathcal{C} = M_{A}^\mathrm{T}\chi M_{B} \;  \label{eq:corre_matrix_bloch}
\end{align}
where $M_{A}$ and $M_{B}$ are the Bloch representations of $\boldsymbol{A}$ and $\boldsymbol{B}$ defined in \cref{eq:measure_vec_bloch}. With this framework, one may observe that the traceless part of the measurement $\boldsymbol{A}$ can be interpreted as the vertices of a polygon or polyhedron in $(d^2-1)$-dimensional space, which provides valuable insights for the further analysis of \cref{thm:sep_cri}.

Let us consider the simplest scenario where $t_{\mu}^{A}=0$, $\vec{a}_{1}=\vec{0}$, and $\vec{a}_{\mu}=(0,\cdots,0,1,0,\cdots,0)^{\mathrm{T}}$ for $\mu=2,\cdots,d_{A}^2$. Here, all components of $\vec{a}_{\mu}$ are zero except for the $(\mu-1)$th component which equals 1. It is straightforward to determine that $\beta_A=\frac{2(d_A-1)}{d_A}$ (analogously for $\beta_B$). The separability condition can be expressed as:
\begin{align}
\|\mathcal{C}\|_{\mathrm{tr}} = \|\mathcal{T}\|_{\mathrm{tr}} \leq \sqrt{\frac{4(d_{A}-1)(d_{B}-1)}{d_{A}d_{B}}} \; .
\label{eq:vicente_cri}
\end{align}
Notably, Eq.(\ref{eq:vicente_cri}) simply reproduces the de Vicente's correlation matrix criterion \cite{vicente07}. 

Furthermore, by setting \( t_{1}^{A} = \sqrt{2d_A} h_{A} \), \( t_{1}^{B} = \sqrt{2d_B} h_{B} \), and \( t_{\mu}^{A} = t_{\mu}^{B} = 0 \) for \( \mu \neq 1 \), and choosing \( \{\vec{a}_{\mu}\} \) and \( \{\vec{b}_{\mu}\} \) as previously defined, we obtain
\[
\|\mathcal{C}\|_{\mathrm{tr}} \leq \sqrt{\frac{2(d_{A}-1+h_{A}^2)}{d_{A}}} \sqrt{\frac{2(d_{B}-1+h_{B}^2)}{d_{B}}},
\]
which is precisely the Sarbicki \emph{et al.}'s criterion \cite{sarbicki20}. Additionally, if \( \boldsymbol{A} \) (and similarly for \( \boldsymbol{B} \)) has the same trace and its traceless part corresponds to a regular simplex, specifically,
\begin{align}
M_{A} = \left[
\begin{matrix}
\frac{t_{A}}{\sqrt{2d_{A}}} & \frac{t_{A}}{\sqrt{2d_{A}}} & \cdots & \frac{t_{A}}{\sqrt{2d_{A}}} \\
\vec{a}_{1} & \vec{a}_{2} & \cdots & \vec{a}_{d_{A}^2}
\end{matrix}
\right] \; ,
\end{align}
where \( \{\vec{a}_{\mu}\} \) forms a regular \( (d_{A}^2-1) \)-simplex, i.e., \( \vec{a}_{\mu} \cdot \vec{a}_{\nu} = \frac{d_{A}^2 \delta_{\mu\nu} - 1}{d_{A}^2 - 1} \) for \( \mu, \nu = 1, \ldots, d_{A}^2 \), a family of separability criteria can be readily derived. The proof and the construction of the regular simplex vertex coordinates are provided in Appendices \ref{appen:proof_obser_regular} and \ref{appen:regular_simplex}.
\begin{observation}\label{obser:simplex_cri}
Given $\boldsymbol{A}$ and $\boldsymbol{B}$, let their trace parts satisfy $t_{\mu}^{A}=t_{A},t_{\mu}^{B}=t_{B}$ and traceless parts correspond respectively to a regular simplex. The correlation matrix of separable state $\rho$ then satisfies
\begin{align}
\|\mathcal{C}\|_{\mathrm{tr}} \leq \sqrt{t_{A}^2+\frac{2d_{A}}{d_{A}+1}}\sqrt{t_{B}^2+\frac{2d_{B}}{d_{B}+1}} \; ,
\label{eq:simplex_cri}
\end{align}
otherwise, $\rho$ is entangled. Here, parameters $t_{A}$ and $t_{B}$ are arbitrary real numbers.
\end{observation}

The \cref{obser:simplex_cri} represents a family of separability conditions for various values of $t_{A}$ and $t_{B}$, with many existing criteria appearing as particular cases of this family. Specifically, when $t_{X}$ takes the values $\sqrt{\frac{2d_{X}}{d_{X}^2-1}}$ and $\sqrt{\frac{2d_{X}}{d_{X}-1}}$ ($X=A,B$), it yields the well-known CCNR criterion \cite{rudolph03,chen03} and the entanglement criterion via SIC-POVM (ESIC) \cite{shang18}, respectively. For clarity and convenience, we have summarized these results in \cref{tab:summary_criteria}.
\begin{table}[ht]
\caption{\label{tab:summary_criteria}Here, we summarize the relations between our criteria and these mentioned in the main text without local filtering transformation, in which CCNR and ESIC are derived from \cref{obser:simplex_cri} with $t=\sqrt{\frac{2d}{d^2-1}}$ and $t=\sqrt{\frac{2d}{d-1}}$, respectively.}
\begin{tabular*}{\hsize}{@{}@{\extracolsep{\fill}}llll@{}}
\hline\hline 
& \\[-8pt]  %可以避免文字偏上来调整文字与上边界的距离
Criteria & \hspace{-0.8em}$M_A/M_B$ & $\kappa_A\kappa_B$ \\ 
\hline
& \\[-8pt]  %可以避免文字偏上来调整文字与上边界的距离
Vicente \cite{vicente07} & \hspace{-1.2em}$\left[
\begin{smallmatrix}
0 & \vec{0}^{\mathrm{T}} \\
\vec{0} & \mathds{1}_{d^2-1}
\end{smallmatrix}
\right]$ & \hspace{-1.2em}$\sqrt{\frac{4(d_{A}-1)(d_{B}-1)}{d_{A}d_{B}}}$ \\ 
% \midrule
& \\[-8pt]  %可以避免文字偏上来调整文字与上边界的距离
Sarbicki \cite{sarbicki20} & \hspace{-1.2em}$\left[
\begin{smallmatrix}
h & \vec{0}^{\mathrm{T}} \\
\vec{0} & \mathds{1}_{d^2-1}
\end{smallmatrix}
\right]$ & \hspace{-3em}$\sqrt{\frac{4(d_{A}-1+h_{A}^2)(d_{B}-1+h_{B}^2)}{d_{A}d_{B}}}$ \\
% \midrule
& \\[-8pt]  %可以避免文字偏上来调整文字与上边界的距离
This work & \hspace{-1.2em}$\left[
\begin{smallmatrix}
\frac{t}{\sqrt{2d}} & \cdots & \frac{t}{\sqrt{2d}} \\
\vec{a}_{1} & \cdots & \vec{a}_{d^2}
\end{smallmatrix}
\right]$ & \hspace{-3.2em}$\sqrt{t_{A}^2+\frac{2d_{A}}{d_{A}+1}}\sqrt{t_{B}^2+\frac{2d_{B}}{d_{B}+1}}$ \\
% \midrule 
& \\[-8pt]  %可以避免文字偏上来调整文字与上边界的距离
CCNR \!\cite{rudolph03,chen03} & \hspace{-1.2em}$\left[
\begin{smallmatrix}
\frac{1}{\sqrt{d^2-1}} & \cdots & \frac{1}{\sqrt{d^2-1}} \\
\vec{a}_{1} & \cdots & \vec{a}_{d^2}
\end{smallmatrix}
\right]$ & \hspace{-0.5em}$\frac{2d_{A}d_{B}}{\sqrt{(d_{A}^2-1)(d_{B}^2-1)}}$ \\ 
% \midrule
& \\[-8pt]  %可以避免文字偏上来调整文字与上边界的距离 
ESIC \cite{shang18} & \hspace{-1.2em}$\left[
\begin{smallmatrix}
\frac{1}{\sqrt{d-1}} & \cdots & \frac{1}{\sqrt{d-1}} \\
\vec{a}_{1} & \cdots & \vec{a}_{d^2}
\end{smallmatrix}
\right]$ & \hspace{-0.5em}$\frac{4d_{A}d_{B}}{\sqrt{(d_{A}^2-1)(d_{B}^2-1)}}$ \\
& \\[-8pt]  %可以避免文字偏上来调整文字与上边界的距离
\hline\hline 
\end{tabular*}
\end{table}

To elaborate further, we compare \cref{obser:simplex_cri} with several symbolic criteria under specific quantum states: the Horodecki's bound entangled states with white noise $\rho_{\mathrm{H}}(s,p)$, unextendible product bases (UPB) bound entangled states with white noise $\rho_{\mathrm{UPB}}(p)$, and chessboard states $\rho_{\mathrm{CB}}$. The exact forms of these quantum states are shown in the \cref{appen:bound_ent_states}.

\begin{figure*}[ht]
% \centering
\subfloat[Horodecki's bound entangled states with white noise]{\label{fig:horodecki_bound}\includegraphics[width=0.5\textwidth]{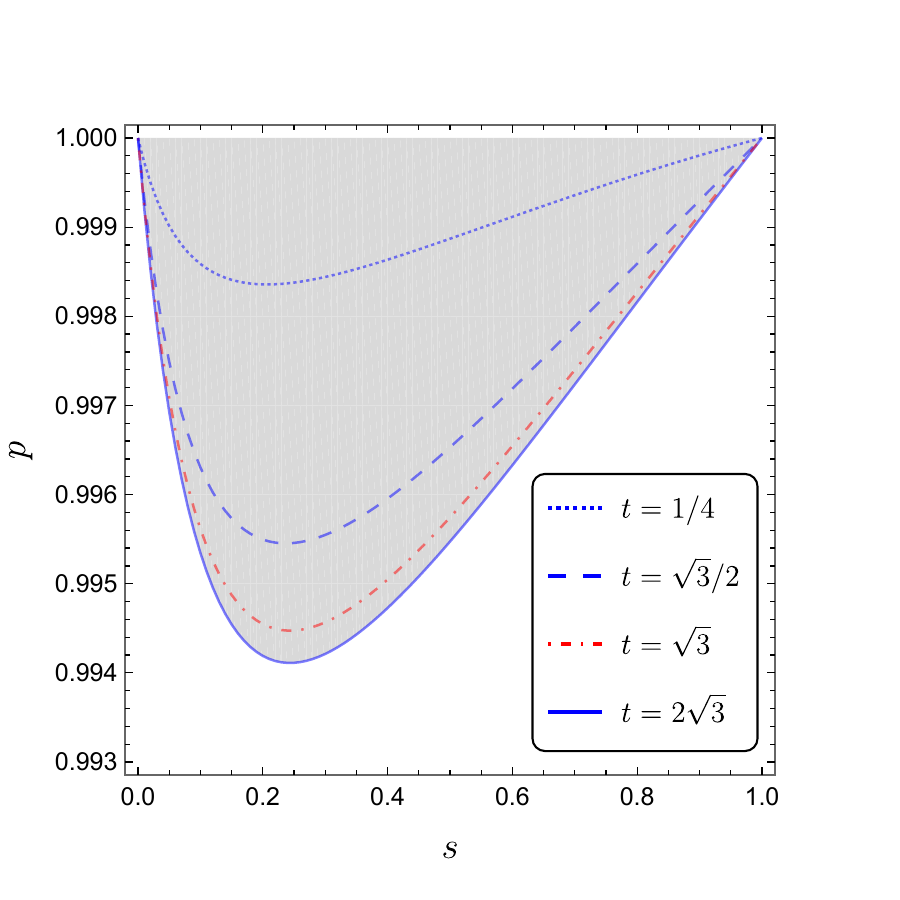}}
\subfloat[UPB and chessboard states]{\label{fig:upb_chessboard}\includegraphics[width=0.5\textwidth]{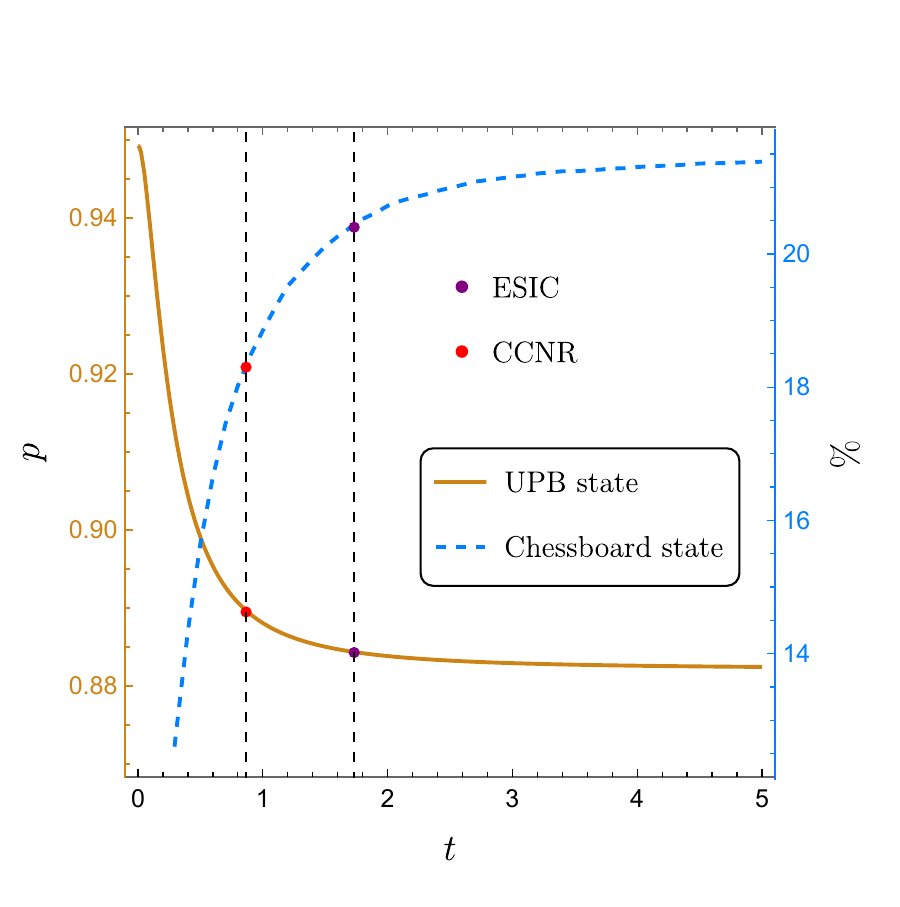}}
\caption{Entanglement detection for three classes of bound entangled states. (a) Horodecki's $3\times 3$ bound entangled states with white noise $\rho_{\mathrm{H}}(s,p)$. (b) UPB bound entangled states with white noise $\rho_{\mathrm{UPB}}(p)$ and $50000$ randomly generated chessboard states $\rho_{\mathrm{CB}}$. The larger $t$ detects the more entangled states as shown in (a) and (b), in which $t=\sqrt{3}/2$ and $t=\sqrt{3}$ correspond to CCNR and ESIC criteria, respectively. Diagrams indicate explicitly that the criterion given in Observation 1 outperform CCNR and ESIC criteria under the three typical classes of bound entangled states.}
\label{fig:bound_ent_exp}
\end{figure*}

In \cref{fig:bound_ent_exp}, we compare the performance of \cref{obser:simplex_cri}, the CCNR criterion, and the ESIC criterion for three classes of bound entangled states. Here, a larger value of $t$ indicates greater robustness against white noise, enabling the detection of a larger portion of entangled states. \cref{fig:bound_ent_exp}(a) presents results for Horodecki’s bound entangled states with white noise, while \cref{fig:bound_ent_exp}(b) details the detection results for the UPB state with white noise (solid peru line) and $50000$ randomly generated chessboard states (blue dashed line). Notably, as illustrated in the figure, \cref{obser:simplex_cri} reproduces the CCNR and ESIC criteria as two special cases when $t=\sqrt{3}/2$ and $t=\sqrt{3}$, respectively.

When the two involved subsystems share the same dimension, an enhanced separability condition is readily established (see \cref{appen:proof_obser_regular_enhance} for the proof):
\begin{observation}\label{obser:simplex_cri_tr}
Given $\boldsymbol{A}=\boldsymbol{B}$, let their trace parts satisfy $t_{\mu}^{A}=t_{\mu}^{B}=t$ and traceless parts correspond to a regular simplex. The correlation matrix of the $d\times d$ separable state $\rho$ then satisfies
\begin{align}
\|\mathcal{C}\|_{\mathrm{tr}} \leq t^2+\frac{2d}{d+1} \; ,
\label{eq:simplex_cri_tr_norm}
\end{align}
and 
\begin{align}
\tr[\mathcal{C}] \geq t^2 - \frac{2d}{d^2-1}\; ,
\label{eq:simplex_cri_tr}
\end{align}
otherwise, $\rho$ is entangled. Here, $t$ is arbitrary real parameter.
\end{observation}

\begin{figure}[htb]
\centering
\includegraphics[width=0.45\textwidth]{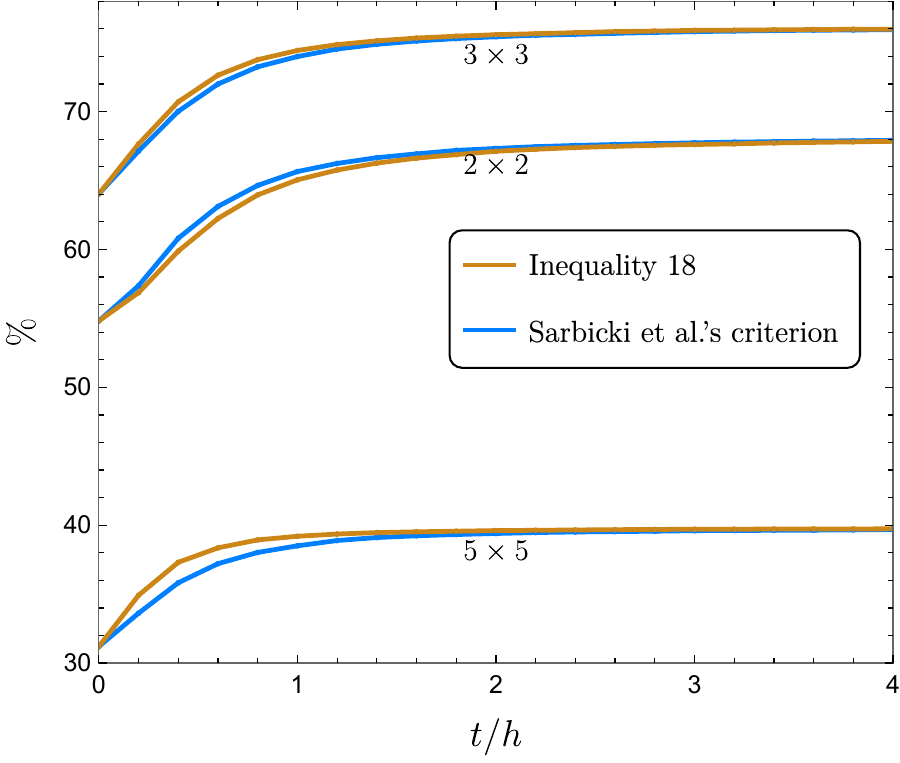}
\caption{Entanglement detection for the randomly generated $2\times 2$, $3\times 3$ and $5\times 5$ states with the Hilbert-Schmidt distribution. With the parameters $t$ or $h$ increasing, the two inequalities tend to be agreeable and detect the same fractions. }
\label{fig:random_complex_gaussian}
\end{figure}

We now compare the criterion proposed by Sarbicki \emph{et al.} \cite{sarbicki20} with \cref{obser:simplex_cri_tr}. To this end, we first consider the \emph{random $2\times 2$, $3\times 3$ and $5\times 5$ states with the Hilbert-Schmidt distribution}\footnote{If a square random matrix $A$ whose entries are drawn independently according to the standard complex Gaussian distribution, then the density matrix $\rho = AA^{\dagger}/\tr[AA^{\dagger}]$ obeys the Hilbert-Schmidt distribution \cite{shang18}.}. In \cref{fig:random_complex_gaussian}, we plot the fractions of entangled states identified by \cref{eq:simplex_cri_tr_norm} (peru solid line) and the criterion proposed by Sarbicki \emph{et al.} (blue solid line) for 50000 randomly generated $2\times 2$, $3\times 3$ and $5\times 5$ states following the Hilbert-Schmidt distribution. As the parameters $t$ and $h$ increase, the two inequalities tend to align, detecting the same fractions of entangled states. Therefore, we conjecture that the criterion proposed by Sarbicki \emph{et al.} and \cref{obser:simplex_cri} are equivalent when the entire parameter spaces are explored. Notably, a significantly larger percentage of entangled states is detected in the $3\times 3$ case compared to other dimensions, as highlighted in \cite{shang18}, underscoring the uniqueness of dimension $3$.

Next, we specifically consider arbitrary dimensional Werner states \cite{werner89,yang21} 
\begin{align}
\rho_{\mathrm{W}}\! = \!\frac{1}{d^2} \mathds{1}\otimes \mathds{1}\! + \!\frac{1}{4} \sum_{\mu=1}^{d^2-1} \frac{2(d\phi-1)}{d(d^2-1)} \pi_{\mu} \otimes \pi_{\mu} \; , 
\label{eq:bloch_werner}
\end{align}
where the parameter range is $\phi\in[-1,1]$. The Werner states are separable if and only if $\phi\geq 0$ \cite{werner89}. For the arbitrary dimensional Werner states, Sarbicki \emph{et al.}'s criteria give $|d\phi-1|\leq\sqrt{d-1+h_A^2}\sqrt{d-1+h_B^2}$, if $\rho_{\mathrm{W}}$ is separable. The optimal inequality constraint is achieved by  $\min_{\{h_A,h_B\}}\sqrt{d-1+h_A^2}\sqrt{d-1+h_B^2}=d-1$, which recognizes the entangled range $\phi\in[-1,-\frac{d-2}{d})$. Notably, the intervals of entanglement detectable by Sarbicki \emph{et al.}'s criteria tend to diminish to zero as the dimensionality increases. A straightforward calculation demonstrates that \cref{eq:simplex_cri_tr} serves as a necessary and sufficient condition applicable across the entire parameter space $\phi\in[-1, 0)$ for the arbitrary dimensional Werner states.

\section{Enhanced separability criteria and local filtering transformation}

Encouraged by the results in Fig.\ref{fig:bound_ent_exp}, one might naturally ask whether a larger $t$ indicates a more powerful criterion. Specifically, the question arises as to whether the ESIC criterion is stronger than the CCNR criterion, a conjecture proposed in Ref. \cite{shang18}. Unfortunately, \cref{eq:simplex_cri} cannot tell more, since the left hand side (LHS) of it contains as well parameters $t_{A}$ and $t_{B}$.

\begin{figure}
\centering
\includegraphics[width=0.5\textwidth]{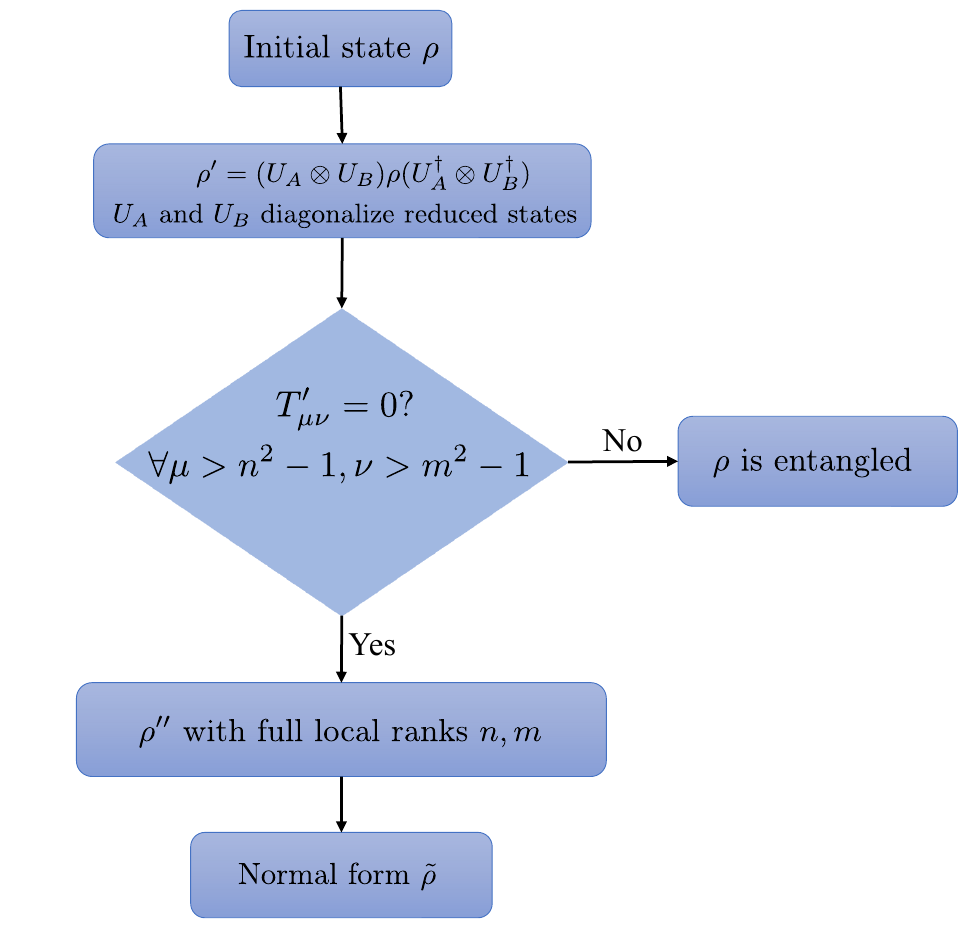}
\caption{The flow diagram of entanglement detection. The bipartite separability problem reduces to the bipartite states with full local ranks and further normal form states.}
\label{fig:flow_diagram_ent_detection}
\end{figure}

Nevertheless, it is found that this question can be readily addressed under the local filtering transformation (LFT), which preserves the separability of a given state, i.e. $\rho\xmapsto{} \tilde{\rho}=(F_{A}\otimes F_{B})\rho(F_{A}\otimes F_{B})^{\dagger}/\tr[(F_{A}\otimes F_{B})\rho(F_{A}\otimes F_{B})^{\dagger}]$ with $F_{A/B}$ the arbitrary invertible matrices \cite{verstraete03}. LFT transforms the full local rank state into a normal form with maximally mixed subsystems \cite{verstraete03,li18-frAQB}
\begin{align}
\tilde{\rho} = \frac{1}{d_{A}d_{B}}\mathds{1}\otimes\mathds{1} + \frac{1}{4}\sum_{\mu,\nu}\widetilde{\mathcal{T}}_{\mu\nu}\pi_{\mu}^{A}\otimes\pi_{\nu}^{B} \; .
\label{eq:normal_form}
\end{align}
It is proved that all $d_A\times d_B$ states with local ranks $n\leq d_A$ and $m\leq d_B$ are either reducible to $n\times m$ bipartite states with full local ranks, or entangled in Ref. \cite{li18-frAQB}. Therefore we only need to focus on the separability problem of quantum states with full local ranks, which has been depicted in the flow diagram of \cref{fig:flow_diagram_ent_detection}.
In this sense, the bipartite separability problem can be addressed through quantum states expressed in the form of \cref{eq:normal_form}. Specifically, we present the following theorem (see \cref{appen:proof_thm_lft} for the proof):
\begin{theorem}\label{thm:sep_cri_lft}
Let measurement vectors $\boldsymbol{A}$ and $\boldsymbol{B}$ satisfy $\sum_{\mu}t_{\mu}^{A}\vec{a}_{\mu}=\sum_{\mu}t_{\mu}^{B}\vec{b}_{\mu}=0$. If a quantum state $\rho$ is separable, one may find the following relation
\begin{align}
\tr\left[\sqrt{\mathcal{A}\widetilde{\mathcal{T}}\mathcal{B}\widetilde{\mathcal{T}}^{\mathrm{T}}}\right] \leq \kappa_{A}\kappa_{B} - \frac{|\vec{t}^{A}||\vec{t}^{B}|}{d_{A}d_{B}} \; ,
\label{eq:gene_cri_lft}
\end{align}
where $\mathcal{A}=\sum_{\mu}\vec{a}_{\mu}\vec{a}_{\mu}^{\mathrm{T}}$ and $\mathcal{B}=\sum_{\mu}\vec{b}_{\mu}\vec{b}_{\mu}^{\mathrm{T}}$.
\end{theorem}
As a result of LFT, \cref{thm:sep_cri_lft} formulates a family of even stronger separability criteria than \cref{thm:sep_cri}. If $\{\vec{a}_{\mu}\}$ constitutes a regular $(d_{A}^2-1)$-simplex, then we readily have  $\sum_{\mu}\vec{a}_{\mu}\vec{a}_{\mu}^{\mathrm{T}}=\frac{d_{A}^2}{d_{A}^2-1}\mathds{1}$ and the following Observation:
\begin{observation}\label{obser:simplex_cri_lft}
Given measurement vectors $\boldsymbol{A}$ and $\boldsymbol{B}$, let their trace parts satisfy $t_{\mu}^{A}=t_{A},t_{\mu}^{B}=t_{B}$,
and traceless parts correspond respectively to a regular simplex. If a quantum state $\rho$ is separable, then in its normal form $\tilde{\rho}$
\begin{align}
\|\widetilde{\mathcal{T}}\|_{\mathrm{tr}} \leq \kappa(t_{A},t_{B}) \; .
\label{eq:simplex_cri_lft}
\end{align}
Here,
\begin{widetext}
\begin{align}
\kappa(t_{A},t_{B}) = \frac{\sqrt{(d_{A}^2-1)(d_{B}^2-1)}}{d_{A}d_{B}}\left(\sqrt{t_{A}^2+\frac{2d_{A}}{d_{A}+1}}\sqrt{t_{B}^2+\frac{2d_{B}}{d_{B}+1}}-|t_{A}t_{B}|\right) \; .
\end{align}
\end{widetext}
\end{observation}

Now that the LHS of \cref{eq:simplex_cri_lft} is independent of parameters $t_{A},t_{B}$, we can strictly compare it with other criteria for different values $t_{A}$ and $t_{B}$. Notice that the minimal value of $\kappa(t_{A},t_{B})$, i.e. $\min_{\{t_{A},t_{B}\}}\kappa(t_{A},t_{B})=2\sqrt{(d_{A}-1)(d_{B}-1)/d_{A}d_{B}}$,
corresponds to the strongest constraint on separability, which reproduces the Proposition 6 of Ref. \cite{guhne07} when $d_{A}=d_{B}$. After LFT, the CCNR and ESIC criteria correspond to $\kappa(t_{A},t_{B})=2-2/\sqrt{d_{A}d_{B}}$ and $4-2\sqrt{(d_{A}+1)(d_{B}+1)/d_{A}d_{B}}$ respectively. This indicates that in normal form, the ESIC criterion is stronger than CCNR, with both coinciding when $d_{A}=d_{B}$. In \cref{tab:summary_criteria_lft}, we present a summary of how to obtain de Vicente’s criterion, Sarbicki et al.’s criterion, CCNR, ESIC, and Proposition 6 in \cite{guhne07} after applying local filtering transformations, highlighting the explicit relationships among them.
\begin{table*}[ht]
\caption{\label{tab:summary_criteria_lft}Here, we summarize the relations between our criteria and all these mentioned in the main text after applying LFT. After applying LFT, for the optimal parameter values, \cref{obser:simplex_cri} coincides with Sarbicki et al.'s criterion, i.e. $\|\widetilde{\mathcal{T}}\|_{\mathrm{tr}} \leq2\sqrt{(d_{A}-1)(d_{B}-1)/d_{A}d_{B}}$, which reproduces the Proposition 6 of Ref. \cite{guhne07} when $d_{A}=d_{B}$ and de Vicente's criterion. The ESIC criterion is stronger than CCNR and they coincide as $d_{A}=d_{B}$ after applying LFT.}
\begin{tabular*}{\linewidth}{@{}@{\extracolsep{\fill}}ccccc@{}}
\hline\hline 
& \\[-8pt]  %可以避免文字偏上来调整文字与上边界的距离
Criteria & $M_A/M_B$ & $\mathcal{A}/\mathcal{B}$ & $\kappa_A\kappa_B-\frac{|\vec{t}^{A}||\vec{t}^{B}|}{d_{A}d_{B}}$ \\ 
\hline 
& \\[-8pt]  %可以避免文字偏上来调整文字与上边界的距离
de Vicente's criterion & $\left[
\begin{smallmatrix}
0 & \vec{0}^{\mathrm{T}} \\
\vec{0} & \mathds{1}_{d^2-1}
\end{smallmatrix}
\right]$ & $\mathds{1}_{d^2-1}$ & $\sqrt{\frac{4(d_{A}-1)(d_{B}-1)}{d_{A}d_{B}}}$ \\ 
% \midrule
& \\[-8pt]  %可以避免文字偏上来调整文字与上边界的距离
Sarbicki et al.'s criterion & $\left[
\begin{smallmatrix}
h & \vec{0}^{\mathrm{T}} \\
\vec{0} & \mathds{1}_{d^2-1}
\end{smallmatrix}
\right]$ & $\mathds{1}_{d^2-1}$ & $\sqrt{\frac{4(d_{A}-1+h_{A}^2)(d_{B}-1+h_{B}^2)}{d_{A}d_{B}}}-\frac{|h_A||h_B|}{d_Ad_B}$ \\
% \midrule
& \\[-8pt]  %可以避免文字偏上来调整文字与上边界的距离
This work & $\left[
\begin{smallmatrix}
\frac{t}{\sqrt{2d}} & \cdots & \frac{t}{\sqrt{2d}} \\
\vec{a}_{1} & \cdots & \vec{a}_{d^2}
\end{smallmatrix}
\right]$ & $\frac{d^2}{d^2-1}\mathds{1}_{d^2-1}$ & $\sqrt{t_{A}^2+\frac{2d_{A}}{d_{A}+1}}\sqrt{t_{B}^2+\frac{2d_{B}}{d_{B}+1}}-|t_A||t_B|$ \\
% \midrule 
& \\[-8pt]  %可以避免文字偏上来调整文字与上边界的距离
CCNR & $\left[
\begin{smallmatrix}
\frac{1}{\sqrt{d^2-1}} & \cdots & \frac{1}{\sqrt{d^2-1}} \\
\vec{a}_{1} & \cdots & \vec{a}_{d^2}
\end{smallmatrix}
\right]$ & $\frac{d^2}{d^2-1}\mathds{1}_{d^2-1}$ & $\frac{2(d_{A}d_{B}-\sqrt{d_Ad_B})}{\sqrt{(d_{A}^2-1)(d_{B}^2-1)}}$ \\ 
% \midrule 
& \\[-8pt]  %可以避免文字偏上来调整文字与上边界的距离
ESIC & $\left[
\begin{smallmatrix}
\frac{1}{\sqrt{d-1}} & \cdots & \frac{1}{\sqrt{d-1}} \\
\vec{a}_{1} & \cdots & \vec{a}_{d^2}
\end{smallmatrix}
\right]$ & $\frac{d^2}{d^2-1}\mathds{1}_{d^2-1}$ & $\frac{4d_{A}d_{B}-2\sqrt{d_Ad_B(d_A+1)(d_B+1)}}{\sqrt{(d_{A}^2-1)(d_{B}^2-1)}}$ \\
& \\[-8pt]  %可以避免文字偏上来调整文字与上边界的距离
Proposition 6 in \cite{guhne07} & $\left[
\begin{smallmatrix}
0 & \cdots & 0 \\
\vec{a}_{1} & \cdots & \vec{a}_{d^2}
\end{smallmatrix}
\right]$ & $\frac{d^2}{d^2-1}\mathds{1}_{d^2-1}$ & $\sqrt{\frac{4d_Ad_B}{(d_A+1)(d_B+1)}}$ \\ 
& \\[-8pt]  %可以避免文字偏上来调整文字与上边界的距离
\hline\hline 
\end{tabular*}
\end{table*}

\section{Relations to entanglement witness}

An entanglement witness is a Hermitian operator which has nonnegative expectation values for all separable states and negative one for some entangled states \cite{horodecki01,terhal02}. Entanglement witness provides an experimental friendly tool to analyse entanglement \cite{guhne09}. We next reformulate the \cref{thm:sep_cri} in terms of entanglement witness. To this end, we provide a new understanding for the trace norm of correlation matrix for observables $m$-tuple $\boldsymbol{A}$ and $n$-tuple $\boldsymbol{B}$. Similar to the orbit of a set under group action \cite{dummit03}, one can define the orbit of $m$-tuple $\boldsymbol{A}$ under the orthogonal group $O(m)$
\begin{align}
\mathcal{O}(\boldsymbol{A}) = \{O\boldsymbol{A}|O\in O(m)\} \; .
\end{align}
Here, $\boldsymbol{A}'\in \mathcal{O}(\boldsymbol{A})$ implies $A'_{\mu}=\sum_{\nu}O_{\mu\nu}A_{\nu},\mu=1,\cdots,m$. It is readily to prove that the all elements of $\mathcal{O}(\boldsymbol{A})$ yield the same MIBS up to a rotation. We can define the orbit $\mathcal{O}(\boldsymbol{B})$ of $n$-tuple $\boldsymbol{B}$ as well. Naturally, given $m$-tuple $\boldsymbol{A}$ and $n$-tuple $\boldsymbol{B}$, the orbit of the correlation matrix is defined as 
\begin{align}
\mathcal{O}(\mathcal{C})\! \equiv\! \{\braket{\!A_\mu\otimes B_\nu\!}|\boldsymbol{A}\!\in\! \mathcal{O}(\boldsymbol{A}),\boldsymbol{B}\!\in \!\mathcal{O}(\boldsymbol{B})\!\} \; .
\end{align}
Without loss of generality, assume $m\geq n$.
By padding $\boldsymbol{B}$ with zero matrices, we can reshape it into an $m$-tuple. Consequently, the trace norm of the correlation matrix can be reformulated as follows (see \cref{appen:proof_svd_corre_tr} for proof)
\begin{align}
\|\mathcal{C}\|_{\mathrm{tr}} = \max_{\mathcal{C}\in\mathcal{O}(\mathcal{C})}\tr[\mathcal{C}] \; .
\label{eq:corre_tr}
\end{align}
The von Neumann's trace theorem \cite{horn13} tells us that the maximum is reached when $\boldsymbol{A}'=P\boldsymbol{A}$ and $\boldsymbol{B}'=Q\boldsymbol{B}$, where $P,Q$ achieve a singular value decomposition (SVD) of $\mathcal{C}$, i.e. $\mathcal{C}=P\Sigma Q^{\mathrm{T}}$. Then, we have
\begin{align}
\|\mathcal{C}\|_{\mathrm{tr}} = \max_{\mathcal{C}\in\mathcal{O}(\mathcal{C})}\tr[\mathcal{C}] = \sum_{\mu=1}^{r}\braket{A'_{\mu}\otimes B'_{\mu}}\; .
\label{eq:svd_corre_tr}
\end{align}
Here, $r=\operatorname{rank}(\mathcal{C})$. \cref{eq:svd_corre_tr} shows that there exist the optimal observables $A'_\mu$ and $B'_\mu$ to witness entanglement. Thus, one may immediately get an inequality of correlation function similar to CHSH inequality \cref{eq:chsh} for separable state
\begin{align}
\sum_{\mu=1}^{r}\braket{A'_{\mu}\otimes B'_{\mu}} \leq \kappa \; ,
\label{eq:corr_func_sep}
\end{align}
where $\kappa=\kappa_{A}\kappa_{B}$. Obviously, \cref{eq:corr_func_sep} can be reformulated as the following entanglement witness
\begin{align}
\mathcal{W} = \kappa\mathds{1}\otimes\mathds{1} - \sum_{\mu=1}^{r}A'_{\mu}\otimes B'_{\mu} \; .
\label{eq:ent_witness_opt}
\end{align}
This implies that, based on the extended Bloch space, entanglement witnesses can be easily generated for generic observables.

\section{Concluding remarks}

In the framework of measurement-induced Bloch space, by dint of a novel scheme we obtain a family of stronger separability criteria for entangled system. These criteria encompass many of the well-known existing results, meanwhile explicitly show why they work, i.e. the correlation matrix of a separable state is constrained by the measurement-induced Bloch space. Moreover, the new criteria can be transformed into entanglement witness operators, which makes the construction of entanglement witness from generic observables even transparent. Finally, It is noteworthy that the method developed in this work is in principle extendable to the multipartite systems by introducing the corresponding multipartite correlation tensors, though some tedious work needs to be done.

\section*{Acknowledgements}

This work was supported in part by the National Natural Science Foundation of China(NSFC) under the Grants 12475087, 12235008, the Fundamental Research Funds for the Central Universities and by the University of Chinese Academy of Sciences, and China Postdoctoral Science Foundation funded project No. 2024M753174.

\bibliographystyle{quantum}
\bibliography{ref.bib}

\onecolumn
\appendix

\section{The proof of Theorem 1}\label{appen:proof_thm_sep}

Given $m$-tuple $\boldsymbol{A}$ and $n$-tuple $\boldsymbol{B}$ with $\sum_{\mu}t_{\mu}^{A}\vec{a}_{\mu}=\sum_{\mu}t_{\mu}^{B}\vec{b}_{\mu}=0$, the correlation matrices of the separable states can be expressed as 
\begin{align}
\mathcal{C} = \sum_{i}p_{i}(\vec{t}\,^{A}/d_{A}+\vec{\alpha}_{i})(\vec{t}\,^{B}/d_{B}+\vec{\beta}_{i})^{\mathrm{T}} \; .
\end{align}
Here, $\vec{\alpha}_{i}$ and $\vec{\beta}_{i}$ are the measurement-induced Bloch vectors (MIBVs) associated with $\boldsymbol{A}$ and $\boldsymbol{B}$ as defined in the main text. Employing the convexity property of trace norm $\|X\|_{\operatorname{tr}}=\tr\left[\sqrt{X^{\dagger}X}\right]$, we have 
\begin{align}
\|\mathcal{C}\|_{\mathrm{tr}} &\leq \sum_{i}p_{i}|\vec{t}\,^{A}/d_{A}+\vec{\alpha}_{i}||\vec{t}\,^{B}/d_{B}+\vec{\beta}_{i}| \\ 
&= \sqrt{(\vec{t}\,^{A})^2/d_{A}^2+\vec{\alpha}_{i}^2}\sqrt{(\vec{t}\,^{B})^2/d_{B}^2+\vec{\beta}_{i}^2} \; .
\end{align}
Here, we make use of $\sum_{\mu}t_{\mu}^{A}\vec{a}_{\mu}=\sum_{\mu}t_{\mu}^{B}\vec{b}_{\mu}=0$ and $|\cdot|$ denotes Euclidean norm of a real vector. So, the correlation matrices of the separable states are limited by the size of the measurement-induced Bloch space (MIBS). And we have defined the parameter $\beta_A$
\begin{align}
% \mathcolorbox{yellow}{
\beta_A=\sup_{\rho}\sum_{k=1}^{m}\braket{\bar{A}_k}_{\rho}^2=\sup_{\vec{\alpha}\in \boldsymbol{B}(A)}|\vec{\alpha}|^2
\end{align}
similarly for $\beta_B$, which yields
\begin{align}
\|\mathcal{C}\|_{\operatorname{tr}} \leq \sqrt{(\vec{t}\,^{A})^2/d_{A}^2+\beta_A}\sqrt{(\vec{t}\,^{A})^2/d_{A}^2+\beta_B} \; .
\end{align}
Next, we show that how to obtain an upper bound of $\beta_{A/B}$. Given an $m$-tuple $\boldsymbol{A}$, quantum states can be transformed into the measurement-induced Bloch vector (MIBVs) which constitute the MIBS. It is proved that MIBS is a proper subset of the following hyper-ellipsoid 
\begin{align}
\mathcal{B}(\boldsymbol{A}) \subseteq \left\{\vec{\alpha}\Big|\sqrt{\frac{d}{d\tr[\rho^2]-1}}\vec{\alpha}^{\mathrm{T}}\Omega^{-}\sqrt{\frac{d}{d\tr[\rho^2]-1}}\vec{\alpha}\leq 1\right\} \; .
\label{eq:bloch_ellipsoid}
\end{align}
Here, $\Omega^{-}$ denotes the Moore-Penrose inverse of $\Omega$ which is the unique matrix satisfying the following Moore-Penrose conditons \cite{penrose55}
\begin{align}
\Omega\Omega^{-}\Omega = \Omega \; , \; \Omega^{-}\Omega\Omega^{-} = \Omega^{-} \; , \\
(\Omega\Omega^{-})^{\dagger} = \Omega\Omega^{-} \; , \; (\Omega^{-}\Omega)^{\dagger} = \Omega^{-}\Omega \; .
\end{align}
The Moore-Penrose inverse $\Omega^{-}$ coincides with the matrix inverse when $\Omega$ is invertible. Matrix $\Omega$ completely depicts the properties of the ellipsoid defined by \cref{eq:bloch_ellipsoid}, such as the lengths of the semi-axes are given by $\sqrt{\frac{(d\tr[\rho^2]-1)\lambda_{i}}{d}}$, with $\lambda_{i}$ being the eigenvalues of $\Omega$; the eigenvectors of $\Omega$ determines the directions of the semi-axes \cite{boyd06}. Obviously, Bloch space is a special case of $\boldsymbol{A}=\vec{\pi}$. So, for arbitrary $\boldsymbol{A}$, we have 
\begin{align}
\beta_A \leq \frac{d-1}{d}\lambda_{\max}(\Omega) \; ,
\end{align}
where $\lambda_{\max}(X)$ is the maximal eigenvalue of matrix $X$. In particular, we have $\beta_A=\frac{d-1}{d}\lambda_{\max}(\Omega)$, when \cref{eq:bloch_ellipsoid} is a hypersphere.

\section{The proof of Observation 1}\label{appen:proof_obser_regular}

If $\boldsymbol{A}$ has the same trace and the traceless part corresponds to a regular simplex, namely
\begin{align}
M_{A} = \left[
\begin{matrix}
\frac{t_{A}}{\sqrt{2d_{A}}} & \frac{t_{A}}{\sqrt{2d_{A}}} & \cdots & \frac{t_{A}}{\sqrt{2d_{A}}} \\
\vec{a}_{1} & \vec{a}_{2} & \cdots & \vec{a}_{d_{A}^2} 
\end{matrix}
\right] \; ,
\end{align}
where $\{\vec{a}_{\mu}\}$ constitutes a regular $(d_{A}^2-1)$-simplex, i.e. $\left\{\vec{a}_{\mu}\Big|\vec{a}_{\mu}\cdot\vec{a}_{\nu}=\frac{d_{A}^2\delta_{\mu\nu}-1}{d_{A}^2-1}\right\},\mu,\nu=1,\cdots,d_{A}^2$. Then we have $(\Omega_{A})_{\mu\nu}=2\vec{a}_{\mu}\cdot\vec{a}_{\nu}=2\frac{d_{A}^2\delta_{\mu\nu}-1}{d_{A}^2-1}$, whose matrix form is 
\begin{align}
\Omega_{A} = \frac{2}{d_{A}^2-1}\left(d_{A}^2\mathds{1} - J_{d_{A}^2}\right) \; ,
\end{align}
where, $J_{d_{A}^2}$ is a $d_{A}^2$-dimensional all-ones matrix all whose elements are equal to one. On account of $J_{d_{A}^2}^2=d_{A}^2J_{d_{A}^2}$, it is readily to find eigenvalues of $J_{d_{A}^2}$, i.e. $d_{A}^2$ and $0$ ($d_{A}^2-1$-fold degeneracy). So, the MIBS $\mathcal{B}(\boldsymbol{A})$ is a degenerative hypersphere with radius $\sqrt{\frac{2d_A}{d_A+1}}$. Hence, we have $\beta_A=\frac{2d_A}{d_A+1}$, and 
\begin{align}
\kappa_{A} = \sqrt{t_{A}^2+\frac{2d_{A}}{d_{A}+1}} \; ,
\end{align}
similarly for $\kappa_{B}$. So, for the separable state, we have
\begin{align}
\|\mathcal{C}\|_{\mathrm{tr}} \leq \sqrt{t_{A}^2+\frac{2d_{A}}{d_{A}+1}}\sqrt{t_{B}^2+\frac{2d_{B}}{d_{B}+1}} \; .
\end{align}
Here, parameters $t_{A}$ and $t_{B}$ can be arbitrary real numbers. And next, we prove that, when $t_{X}=\sqrt{\frac{2d_{X}}{d_{X}^2-1}}$ and $t_{X}=\sqrt{\frac{2d_{X}}{d_{X}-1}}$ for $X=A,B$, \cref{eq:simplex_cri} reproduces CCNR and ESIC (entanglement criterion via SIC-POVM) criteria, respectively. Firstly, we review the CCNR criterion. A bipartite state can be expressed as 
\begin{align}
\rho=\frac{1}{4}\sum_{\mu,\nu}\chi_{\mu\nu}\Pi_{\mu}^{A}\otimes\Pi_{\nu}^{B} \; .
\end{align}
Here, $\Pi_{0}^{X}=\sqrt{\frac{2}{d_{X}}}\mathds{1}$ and $\{\Pi_{\mu}^{X}=\pi_{\mu}^{X}\}_{\mu=1}^{d_{X}^2-1}$ is generators of $\mathfrak{su}(d_{X})$ Lie algebra for $X=A,B$; $\{\Pi_{\mu}^{X}\}_{\mu=0}^{d_{X}^2-1}$ satisfies the orthogonal relation $\tr[\Pi_{\mu}^{X}\Pi_{\nu}^{X}]=2\delta_{\mu\nu}$. Making use of the Schmidt decomposition in operator space, we have 
\begin{align}
\rho = \frac{1}{4}\sum_{\mu,\nu}\chi_{\mu\nu}\Pi_{\mu}^{A}\otimes\Pi_{\nu}^{B} 
= \sum_{k}\sigma_{k} \widetilde{\Pi}_{k}^{A}\otimes\widetilde{\Pi}_{k}^{B} \; .
\end{align}
where $\sum_{k}\sigma_{k}=\frac{1}{2}\|\chi\|_{\mathrm{tr}}$. $\{\widetilde{\Pi}_{k}^{A}\}$ and $\{\widetilde{\Pi}_{k}^{B}\}$ form orthonormal bases of the operator space. CCNR criterion says that, if $\rho$ is separable, then we have $\sum_{k}\sigma_{k}\leq 1$ \cite{rudolph05,chen03}, that is, $\|\chi\|_{\mathrm{tr}}\leq 2$. If $t_{X}=\sqrt{\frac{2d_{X}}{d_{X}^2-1}}$ for $X=A,B$, then we have 
\begin{align}
M_{A}^{\mathrm{T}}M_{A} &= \left[
\begin{matrix}
\frac{1}{\sqrt{d_{A}^2-1}} & \vec{a}_{1} \\
\frac{1}{\sqrt{d_{A}^2-1}} & \vec{a}_{2} \\ 
\vdots & \vdots \\ 
\frac{1}{\sqrt{d_{A}^2-1}} & \vec{a}_{d_{A}^2} 
\end{matrix}
\right]\left[
\begin{matrix}
\frac{1}{\sqrt{d_{A}^2-1}} & \frac{1}{\sqrt{d_{A}^2-1}} & \cdots & \frac{1}{\sqrt{d_{A}^2-1}} \\
\vec{a}_{1} & \vec{a}_{2} & \cdots & \vec{a}_{d_{A}^2} 
\end{matrix}
\right] = \frac{d_{A}^2}{d_{A}^2-1}\mathds{1} \; .
\end{align}
Here, $\vec{a}_{\mu}\cdot\vec{a}_{\nu}=\frac{d_{A}^2\delta_{\mu\nu}-1}{d_{A}^2-1}$ has been employed. Similarly, we have $M_{A}^{\mathrm{T}}M_{A}= \frac{d_{B}^2}{d_{B}^2-1}\mathds{1}$. So, if $t_{X}=\sqrt{\frac{2d_{X}}{d_{X}^2-1}}$ for $X=A,B$, \cref{eq:simplex_cri} gives
$\|\mathcal{C}\|_{\mathrm{tr}}=\|M_{A}^\mathrm{T}\chi M_{B}\|_{\mathrm{tr}}=\frac{d_{A}d_{B}}{\sqrt{(d_{A}^2-1)(d_{B}^2-1)}}\|\chi\|_{\mathrm{tr}}\leq \frac{2d_{A}d_{B}}{\sqrt{(d_{A}^2-1)(d_{B}^2-1)}}$, that is, $\|\chi\|_{\mathrm{tr}}\leq 2$. 

And then, the ESIC criteria says that if $\rho$ is separable, then the correlation matrix of the normalized SIC-POVM $\{E_{\mu}^{A}\}$ and $\{E_{\mu}^{B}\}$ satisfies \cite{shang18}
\begin{align}
\|\mathcal{P}\|_{\mathrm{tr}} \leq 1 \; ,
\end{align}
where $\mathcal{P}_{\mu\nu}=\braket{E_{\mu}^{A}\otimes E_{\nu}^{B}}$, $E_{\mu}^{X}\geq 0$ and $\tr[E_{\mu}^{X}E_{\nu}^{X}]=\frac{d_{X}\delta_{\mu\nu}+1}{2d_{X}},\mu,\nu=1,2,\cdots,d_{X}^2$ for $X=A,B$. It is noteworthy that the original assumption about the existence of SIC-POVM is not essential as shown in \cite{sarbicki20}. In fact, we can obtain ESIC criterion for arbitrary measurement vector $\boldsymbol{A}$ satisfying $\tr[A_{\mu}A_{\nu}]=\alpha\frac{d\delta_{\mu\nu}+1}{2d},\mu,\nu=1,2,\cdots,d^2$ with nonzero real parameter $\alpha$. If $t_{X}=\sqrt{\frac{2d_{X}}{d_{X}-1}}$ for $X=A,B$, we have $\tr[A_{\mu}A_{\nu}]=\frac{t_{A}^2}{d}+2\vec{a}_{\mu}\cdot\vec{a}_{\nu}=\frac{(2d_{A})^2}{d_{A}^2-1}\frac{d_{A}\delta_{\mu\nu}+1}{2d_{A}}$ and \cref{eq:simplex_cri} gives 
\begin{align}
\|\mathcal{C}\|_{\mathrm{tr}}\leq \frac{4d_{A}d_{B}}{\sqrt{(d_{A}^2-1)(d_{B}^2-1)}} \; ,
\end{align}
which reproduces the ESIC criterion due to $\braket{A_{\mu}\otimes B_{\nu}}=\frac{4d_{A}d_{B}}{\sqrt{(d_{A}^2-1)(d_{B}^2-1)}}\braket{E_{\mu}^{A}\otimes E_{\nu}^{B}}$.

\section{The vertex coordinates for regular simplex}\label{appen:regular_simplex}

There are many methods to generate the vertex coordinates for regular simplex, where we give a simple scheme for the reader's convenience. Given the following $n$-dimensional unit vectors 
\begin{align}
\vec{e}_{k} = (b_{1},b_{2},\cdots,b_{k-1},a_{k},0,\cdots,0) \; ,
\end{align}
and
\begin{align}
a_{k} = \frac{\sqrt{(n+1)(n-k+1)}}{\sqrt{n(n-k+2)}},b_{k}=-\frac{\sqrt{\
n+1}}{\sqrt{n(n-k+1)(n-k+2)}},k=1,2,\cdots,n+1,
\end{align}
then $\{\vec{e}_{1},\vec{e}_{2}\cdots,\vec{e}_{n+1}\}$ constitutes a $n$-dimensional regular simplex. The following is an example for $n=8$,
\begin{align}
\left[
\begin{array}{ccccccccc}
1 & -\frac{1}{8} & -\frac{1}{8} & -\frac{1}{8} & -\frac{1}{8} & -\frac{1}{8} & -\frac{1}{8} & -\frac{1}{8} & -\frac{1}{8} \\
0 & \frac{3 \sqrt{7}}{8} & -\frac{3}{8 \sqrt{7}} & -\frac{3}{8 \sqrt{7}} & -\frac{3}{8 \sqrt{7}} & -\frac{3}{8 \sqrt{7}} & -\frac{3}{8 \sqrt{7}} & -\frac{3}{8 \sqrt{7}} & -\frac{3}{8 \sqrt{7}} \\
0 & 0 & \frac{3}{2}\sqrt{\frac{3}{7}} & -\frac{1}{4}\sqrt{\frac{3}{7}} & -\frac{1}{4}\sqrt{\frac{3}{7}} & -\frac{1}{4}\sqrt{\frac{3}{7}} & -\frac{1}{4}\sqrt{\frac{3}{7}} & -\frac{1}{4}\sqrt{\frac{3}{7}} & -\frac{1}{4}\sqrt{\frac{3}{7}} \\
0 & 0 & 0 & \frac{\sqrt{15}}{4} & -\frac{1}{4}\sqrt{\frac{3}{5}} & -\frac{1}{4}\sqrt{\frac{3}{5}} & -\frac{1}{4}\sqrt{\frac{3}{5}} & -\frac{1}{4}\sqrt{\frac{3}{5}} & -\frac{1}{4}\sqrt{\frac{3}{5}} \\
0 & 0 & 0 & 0 & \frac{3}{\sqrt{10}} & -\frac{3}{4 \sqrt{10}} & -\frac{3}{4 \sqrt{10}} & -\frac{3}{4 \sqrt{10}} & -\frac{3}{4 \sqrt{10}} \\
0 & 0 & 0 & 0 & 0 & \frac{3}{4}\sqrt{\frac{3}{2}} & -\frac{1}{4}\sqrt{\frac{3}{2}} & -\frac{1}{4}\sqrt{\frac{3}{2}} & -\frac{1}{4}\sqrt{\frac{3}{2}} \\
0 & 0 & 0 & 0 & 0 & 0 & \frac{\sqrt{3}}{2} & -\frac{\sqrt{3}}{4} & -\frac{\sqrt{3}}{4} \\
0 & 0 & 0 & 0 & 0 & 0 & 0 & \frac{3}{4} & -\frac{3}{4} \\
\end{array}
\right] \; .
\end{align}

\section{The proof of Observation 2}\label{appen:proof_obser_regular_enhance}

Via \cref{obser:simplex_cri}, one immediately obtain the first inequality for the $d\times d$ system, i.e.
\begin{align}
\|\mathcal{C}\|_{\mathrm{tr}} \leq t^2+\frac{2d}{d+1} \; .
\end{align}
In the main text, we reformulate the correlation matrix of the separable state based on the MIBVs, i.e. \cref{eq:simplex_cri_tr}. When the involved two subsystems share the same dimension and the traceless parts of $\boldsymbol{A}=\boldsymbol{B}$ correspond to a regular simplex with trace parts $t_{\mu}^{A}=t_{\mu}^{B}=t$, we have
\begin{align}
\mathcal{C} = \sum_{i}q_{i}(\vec{t}/d+\vec{\alpha}_{i})(\vec{t}/d+\vec{\beta}_{i})^{\mathrm{T}} \; .
\end{align}
Here, $\vec{t}=(t,t,\cdots,t)^{\mathrm{T}}$ and $\vec{\alpha}_{i},\vec{\beta}_{i}\in\mathcal{B}(\boldsymbol{A})$ with $\alpha_{i\mu}=\vec{r}_i\cdot\vec{a}_{\mu},\beta_{i\mu}=\vec{s}_i\cdot\vec{a}_{\mu}$, where $\vec{r}_i$ and $\vec{s}_i$ are Bloch vectors. And then, we have
\begin{align}
\tr[\mathcal{C}] &= \sum_{i}q_i\left[t^2+\frac{\vec{t}\cdot\vec{\alpha}_i+\vec{t}\cdot\vec{\beta}_i}{d}+\vec{\alpha}_i\cdot\vec{\beta}_i\right] \\ 
&= t^2 + \sum_iq_i\vec{\alpha}_i\cdot\vec{\beta}_i \; , \; (\vec{t}\cdot\vec{\alpha}_i=\vec{t}\cdot\vec{\beta}_i=0) \\ 
&= t^2 + \sum_iq_i(\vec{r}_i\cdot\vec{a}_1,\vec{r}_i\cdot\vec{a}_2,\cdots,\vec{r}_i\cdot\vec{a}_{d^2})\left(
\begin{matrix}
\vec{s}_i\cdot\vec{a}_1 \\ 
\vec{s}_i\cdot\vec{a}_2 \\
\vdots \\
\vec{s}_i\cdot\vec{a}_{d^2}
\end{matrix}
\right) \\ 
&= t^2 + \sum_iq_i\left[\left(
\begin{matrix}
\vec{a}_1^{\mathrm{T}} \\ 
\vec{a}_2^{\mathrm{T}} \\
\vdots \\
\vec{a}_{d^2}^{\mathrm{T}}
\end{matrix}
\right)\vec{r}_i\right]^{\mathrm{T}}\left[\left(
\begin{matrix}
\vec{a}_1^{\mathrm{T}} \\ 
\vec{a}_2^{\mathrm{T}} \\
\vdots \\
\vec{a}_{d^2}^{\mathrm{T}}
\end{matrix}
\right)\vec{s}_i\right] \\ 
&= t^2 + \sum_iq_i\vec{r}_i^{\mathrm{T}}\mathcal{A}\vec{s}_i \; , \; (\mathcal{A}=\sum_{\mu}\vec{a}_{\mu}\vec{a}_{\mu}^{\mathrm{T}}=\frac{d^2}{d^2-1}\mathds{1}_{d^2-1}) \\ 
&= t^2 + \frac{d^2}{d^2-1}\sum_iq_i\vec{r}_i\cdot\vec{s}_i \; .
\end{align}
Due to $0\leq\tr[\rho\rho']\leq 1$ with $\rho=\frac{1}{d}\mathds{1}+\frac{1}{2}\vec{r}_i\cdot\vec{\pi}$ and $\rho'=\frac{1}{d}\mathds{1}+\frac{1}{2}\vec{s}_i\cdot\vec{\pi}$, one can readily find  $-\frac{2}{d}\leq\vec{r}_i\cdot\vec{s}_i\leq\frac{2(d-1)}{d}$ which results in 
\begin{align}
t^2 - \frac{2d}{d^2-1}\leq \tr[\mathcal{C}] \leq t^2 + \frac{2d}{d+1} \; .
\end{align}
Because of $\tr[\mathcal{C}]\leq\|\mathcal{C}\|_{\mathrm{tr}}$, we only need to consider the left hand side of this inequality. In all, we have proved that if a $d\times d$ bipartite state is separable, then it satisfies 
\begin{align}
\|\mathcal{C}\|_{\mathrm{tr}} \leq t^2+\frac{2d}{d+1} \; ,
\end{align}
and 
\begin{align}
\tr[\mathcal{C}] \geq t^2 - \frac{2d}{d^2-1} \; .
\end{align}

\section{Bound entangled states}\label{appen:bound_ent_states}

\emph{Horodecki's bound entangled states with white noise:}
\begin{align}
\rho_{\mathrm{H}}(s,p) = p\rho_{\mathrm{H}}(s) + (1-p)\mathds{1}/9 \; , \; 0\leq p \leq 1 \; .
\end{align}
Here, $\rho_{\mathrm{H}}(t)$ is $3\times 3$ Horodecki's bound entangled states \cite{horodecki97}, i.e. 
\begin{align}
\rho_{\mathrm{H}}(s) = \frac{1}{8s+1}\left[
\begin{matrix}
s & 0 & 0 & 0 & s & 0 & 0 & 0 & s \\
0 & s & 0 & 0 & 0 & 0 & 0 & 0 & 0 \\
0 & 0 & s & 0 & 0 & 0 & 0 & 0 & 0 \\
0 & 0 & 0 & s & 0 & 0 & 0 & 0 & 0 \\
s & 0 & 0 & 0 & s & 0 & 0 & 0 & s \\
0 & 0 & 0 & 0 & 0 & s & 0 & 0 & 0 \\
0 & 0 & 0 & 0 & 0 & 0 & \frac{1}{2}(1+s) & 0 & \frac{1}{2}\sqrt{1-s^{2}} \\
0 & 0 & 0 & 0 & 0 & 0 & 0 & s & 0 \\
s & 0 & 0 & 0 & s & 0 & \frac{1}{2}\sqrt{1-s^{2}} & 0 & \frac{1}{2}(1+s)
\end{matrix}
\right] \; .
\end{align}

\noindent
\emph{Unextendible product bases (UPB) bound entangled states:}
\begin{align}
\rho_{\mathrm{UPB}}(p) = p\rho_{\mathrm{UPB}} + (1-p)\mathds{1}/9 \; , \; 0\leq p \leq 1 \; .
\end{align}
Here, $\rho_{\mathrm{UPB}}$ is bound entangled state constructed by UPB \cite{bennett99}, i.e.
\begin{align}
\rho_{\mathrm{UPB}} = \frac{1}{4}\left(\mathds{1}-\sum_{i=0}^4\ket{\psi_i}\bra{\psi_i}\right) \; ,
\end{align}
and 
\begin{align}
&\ket{\psi_0} = \frac{1}{\sqrt{2}}\ket{0}\left(\ket{0}-\ket{1}\right) \; , \; \ket{\psi_1} = \frac{1}{\sqrt{2}}(\ket{0}-\ket{1})\ket{2} \; , \\ \notag
&\ket{\psi_2} = \frac{1}{\sqrt{2}}\ket{2}(\ket{1}-\ket{2}) \; , \; \ket{\psi_3} = \frac{1}{\sqrt{2}}(\ket{1}-\ket{2})\ket{0} \; , \\ \notag
&\ket{\psi_4} = \frac{1}{3}(\ket{0}+\ket{1}+\ket{2})(\ket{0}+\ket{1}+\ket{2}) \; .
\end{align}

\noindent
\emph{Chessboard states \cite{bruss00}:}
\begin{align}
\rho_{\mathrm{CB}} = \mathcal{N}\sum_{i=1}^4\ket{V_i}\bra{V_i} \; .
\end{align}
Here, $\mathcal{N}=1/\sum_{i}\braket{V_i|V_i}$ is normalization constant and the unnormalized vectors $\ket{V_{i}}$ are
\begin{align}
&\ket{V_{1}} = \ket{m, 0, ac/n ; 0, n, 0 ; 0,0,0} \; , \notag \\ 
&\ket{V_{2}} = \ket{0, a, 0 ; b, 0, c ; 0,0,0} \; , \notag \\  
&\ket{V_{3}} = \ket{n, 0,0 ; 0,-m, 0 ; ad/m, 0,0} \; , \notag \\
&\ket{V_{4}} = \ket{0, b, 0 ;-a, 0,0 ; 0, d, 0} \; .
\end{align}
Chessboard states contain six real parameters $m,n,a,b,c,d$. In the main text, we test Observation 1 by randomly generated $50000$ chessboard states, where the six parameters are drawn independently from the standard normal distribution.

\section{The proof of Theorem 2}\label{appen:proof_thm_lft}

By virtue of Bloch representation, the correlation matrix can be written as $\mathcal{C}=M_{A}^\mathrm{T}\chi M_{B}$, so we have
\begin{align}
\|\mathcal{C}\|_{\mathrm{tr}} = \tr\left[\sqrt{\mathcal{C}^{\mathrm{T}}\mathcal{C}}\right] = \tr\left[\sqrt{M_{B}M_{B}^{\mathrm{T}}\chi^{\mathrm{T}} M_{A}M_{A}^{\mathrm{T}}\chi}\right] \; .
\label{eq:corre_tr_norm}
\end{align}
Given arbitrary measurement vectors $\boldsymbol{A}$ and $\boldsymbol{B}$ with $\sum_{\mu}t_{\mu}^{A}\vec{a}_{\mu}=\sum_{\mu}t_{\mu}^{B}\vec{b}_{\mu}=0$, we have 
\begin{align}
M_{A} = \left[
\begin{matrix}
\frac{t_{1}^{A}}{\sqrt{2d_{A}}} & \frac{t_{2}^{A}}{\sqrt{2d_{A}}} & \cdots \\
\vec{a}_{1} & \vec{a}_{2} & \cdots 
\end{matrix}
\right] \; , \; 
M_{B} = \left[
\begin{matrix}
\frac{t_{1}^{B}}{\sqrt{2d_{B}}} & \frac{t_{2}^{B}}{\sqrt{2d_{B}}} & \cdots \\
\vec{b}_{1} & \vec{b}_{2} & \cdots 
\end{matrix}
\right] \; ,
\end{align}
and 
\begin{align}
M_{A}M_{A}^{\mathrm{T}} = \left[
\begin{matrix}
\frac{(\vec{t}^{A})^2}{2d_{A}} & 0 \\
0 & \mathcal{A}
\end{matrix}
\right] = \frac{(\vec{t}^{A})^2}{2d_{A}}\oplus\mathcal{A} \; ,
\end{align}
where $\mathcal{A}=\sum_{\mu}\vec{a}_{\mu}\vec{a}_{\mu}^{\mathrm{T}}$. Similarly, we have $M_{B}M_{B}^{\mathrm{T}}=\frac{(\vec{t}^{B})^2}{2d_{B}}\oplus\mathcal{B}$ and $\mathcal{B}=\sum_{\mu}\vec{b}_{\mu}\vec{b}_{\mu}^{\mathrm{T}}$. For the normal form states, we have $\chi=\frac{2}{\sqrt{d_{A}d_{B}}}\oplus \widetilde{\mathcal{T}}$. And then,  
\begin{align}
\|\mathcal{C}\|_{\mathrm{tr}} &= \tr\left[\sqrt{M_{B}M_{B}^{\mathrm{T}}\chi^{\mathrm{T}} M_{A}M_{A}^{\mathrm{T}}\chi}\right] \\ 
&= \tr\left[\sqrt{\frac{(\vec{t}^{A})^2(\vec{t}^{B})^2}{d_{A}^2d_{B}^2}\oplus\mathcal{B}\widetilde{\mathcal{T}}^{\mathrm{T}}\mathcal{A}\widetilde{\mathcal{T}}}\right] \\ 
&= \frac{|\vec{t}^{A}||\vec{t}^{B}|}{d_{A}d_{B}} + \tr\left[\sqrt{\mathcal{A}\widetilde{\mathcal{T}}\mathcal{B}\widetilde{\mathcal{T}}^{\mathrm{T}}}\right]
\; .
\end{align}
So, if a quantum state $\rho$ is separable, then we have in its normal form $\tilde{\rho}$
\begin{align}
\tr\left[\sqrt{\mathcal{A}\widetilde{\mathcal{T}}\mathcal{B}\widetilde{\mathcal{T}}^{\mathrm{T}}}\right] \leq \kappa_{A}\kappa_{B} - \frac{|\vec{t}^{A}||\vec{t}^{B}|}{d_{A}d_{B}}  \; .
\end{align}

\section{The proof of \cref{eq:svd_corre_tr}}\label{appen:proof_svd_corre_tr}

Without loss of generality, assume $m\geq n$. We add some zero matrices to reshape $\boldsymbol{B}$ into $m$-tuple, then we have
\begin{align}
\max_{\mathcal{C}\in\mathcal{O}(\mathcal{C})}\tr[\mathcal{C}] =& \max_{\{O^{A},O^{B}\}}\sum_{\mu}\Braket{\sum_{\mu'}O_{\mu\mu'}^{A}A_{\mu'}\otimes \sum_{\nu'}O_{\mu\nu'}^{B}B_{\nu'}} \\
&= \max_{\{O^{A},O^{B}\}}\sum_{\mu}\sum_{\mu'\nu'}O_{\mu\mu'}^{A}O_{\mu\nu'}^{B}\braket{A_{\mu'}\otimes B_{\nu'}} \\
&= \max_{\{O^{A},O^{B}\}}\sum_{\mu}\sum_{\mu'\nu'}O_{\mu\mu'}^{A}\mathcal{C}_{\mu'\nu'}(O^{B})^{\mathrm{T}}_{\nu'\mu} \\
&= \max_{\{O^{A},O^{B}\}}\tr[O^{A}\mathcal{C} (O^{B})^{\mathrm{T}}] \\
&= \|\mathcal{C}\|_{\mathrm{tr}} \; .
\end{align}
In the last line, we have employed von Neumann's trace theorem \cite{horn13}.

\end{document}